\documentclass[pdflatex,sn-aps]{sn-jnl}

\usepackage{graphicx}%
\usepackage{multirow}%
\usepackage{amsmath,amssymb,amsfonts}%
\usepackage{amsthm}%
\usepackage{mathrsfs}%
\usepackage[title]{appendix}%
\usepackage{xcolor}%
\usepackage{textcomp}%
\usepackage{manyfoot}%
\usepackage{booktabs}%
\usepackage{comment}
\usepackage{algorithm}%
\usepackage{algorithmicx}%
\usepackage{algpseudocode}%
\usepackage{listings}%
\usepackage[utf8]{inputenc}
\usepackage{xcolor}
\DeclareUnicodeCharacter{202F}{FIX ME!!!!}

\theoremstyle{thmstyleone}%
\theoremstyle{thmstyletwo}%

\theoremstyle{thmstylethree}%

\begin{document}



\title[Article Title]{Optical Fiber–Waveguide Hybrid Architecture for Mid-Infrared Ring Lasers}


\author[1,2]{\fnm{Tinghui} \sur{An}}\email{anth2022@shanghaitech.edu.cn}

\author*[3]{\fnm{T Toney} \sur{Fernandez}}\email{toney.teddyfernandez@unisa.edu.au}

\author*[1]{\fnm{Yuchen} \sur{Wang}}\email{wangyuchen@siom.ac.cn}

\author[1]{\fnm{Yiguang} \sur{Jiang}}\email{jiangyiguang@siom.ac.cn}

\author[4]{\fnm{Benjamin} \sur{Johnston}}\email{benjamin.johnston@mq.edu.au}

\author[5]{\fnm{Solenn} \sur{Cozic}}\email{solenn.cozic@leverrefluore.com}

\author[5]{\fnm{Marcel} \sur{Poulain}}\email{marcel.poulain@leverrefluore.com}

\author[5]{\fnm{Tiphaine} 
\sur{Rault}}

\author[1]{\fnm{Long} \sur{Zhang}}\email{lzhang@siom.ac.cn}

\author[4]{\fnm{Alex} \sur{Fuerbach}}\email{alex.fuerbach@mq.edu.au}

\affil[1]{\orgdiv{Shanghai Institute of Optics and Fine Mechanics}, \orgname{Chinese Academy of Sciences}, \orgaddress{
\city{Shanghai}, \postcode{201800}, 
\country{China}}}

\affil[2]{
\orgname{ShanghaiTech University}, \orgaddress{
\city{Shanghai}, \postcode{200120}, \
\country{China}}}

\affil[3]{
\orgdiv{STEM}, 
\orgname{University of South Australia}, \orgaddress{
\street{Mawson Lakes Campus}, 
\city{South Australia}, \postcode{5095}, \
\country{Australia}}}

\affil[4]{\orgdiv{School of Mathematical and Physical Sciences}, \orgname{Macquarie University}, 
\orgaddress{
\city{NSW}, \postcode{2109}, 
\country{Australia}}}

\affil[5]{
\orgname{Le Verre Fluoré}, \orgaddress{
\street{1 Rue Gabriel Voisin - Campus KerLann}, 
\city{Bruz}, \postcode{35170}, \
\state{Brittany}, 
\country{France}}}


\abstract{
We report a mid-infrared ring laser via fiber-to-waveguide hybrid integration enabled by directional couplers fabricated in fluoride glasses. Directional couplers were inscribed into fluoride glass using femtosecond laser direct writing, forming low-loss waveguides with a high positive refractive index change. The waveguides were optimized for efficient guidance at a wavelength of 2.7 {\textmu}m. Directional couplers with systematically varied coupling gap and interaction lengths were fabricated and experimentally characterized, demonstrating controllable power splitting ratios spanning from complete transmission (100:0) to balanced coupling (50:50). A directional coupler with a 7 {\textmu}m coupling gap and a 1.5 mm interaction length, providing an 80:20 power splitting ratio, was integrated with a cladding-pumped Er:ZBLAN fiber to realize a ring laser cavity. The resulting device produced stable lasing at 2716.7 nm with an output power of 14.2 mW. This work represents the first realization of a fully fiber-chip integrated mid-infrared laser source and opens new avenues for 3D photonic integration in fluoride glasses, enabling design flexibility previously unattainable in traditional fiber-optics architectures.
}

\keywords{fiber-chip integration, ring laser, positive index change fluoride glass, waveguides}



\maketitle

\section{Introduction}

The mid-infrared (mid-IR) spectral region, spanning roughly 2–5 {\textmu}m, holds significant promise for applications in environmental sensing, spectroscopy, free-space communications, and medical diagnostics, owing to the strong molecular fingerprint absorption features in this range~\cite{Jackson2024, Maria2024}. Besides Quantum Cascade Lasers (QCLs), rare-earth-doped fluoride fiber lasers, particularly those based on rare-earth doped ZBLAN ($ZrF_3–BaF_2–LaF_3–AlF_3– NaF$) fibers, have emerged as the most versatile and powerful light sources in this wavelength region. Their broad transmission window, low phonon energy, and established fabrication processes make ZBLAN fibers particularly suited for high-power, low-loss fiber lasers~\cite{Jobin2022, Poulain2024}.

However, despite the rapid development of mid-IR fiber lasers, progress in photonic integration within this wavelength regime has been severely constrained~\cite{Ari2025,fernandez2024}. Traditional fiber architectures rely heavily on discrete, free-space or fiber-spliced components for coupling, filtering, and feedback. These configurations often result in bulky systems that are prone to misalignment, limited in scalability, and restricted to simple planar designs. Unlike in the near-infrared range—where silica-based planar and 3D lightwave circuits and integrated photonics are well-established—mid-IR systems lack an equivalent glass platform that offers both high index tunability and low propagation loss for monolithic integration with fluoride fiber devices.
Nonetheless, significant progress has been made in fiber-based integration, particularly by the Martin Rochette group at McGill University, Canada, who have demonstrated, for the first time to our knowledge, an all-fiber mid-IR ring cavity laser~\cite{Karampour2025}. This achievement was enabled by the development of key components such as a low-loss single-mode ZBLAN fiber coupler, a tapered pump combiner, and an inline polarization controller, all integrated within an Er:ZBLAN fiber to produce stable continuous-wave emission at 2.8$\mu$m with output powers reaching 36 mW. Very recently, the Francesco Prudenzano group from Politecnico di Bari has theoretically designed and experimentally demonstrated a 2×2 optical fiber coupler that operates at a wavelength of 5.0 $\mu$m. The device was realized by employing two single-mode indium fluoride optical fibers enclosed within a low-refractiveindex indium fluoride capillary. The paper underscores the importance of such couplers for mid-infrared applications, particularly in enabling in-line fiber lasers and amplifiers, ring cavities, filtering, and wavelength multiplexing for spectroscopy~\cite{Anelli2025}. In parallel, the group of Martin Bernier at Université Laval, Canada, has pushed the boundaries of mid-IR performance by demonstrating a monolithic all-fiber laser emitting at 3.79 $\mu$m with a record output power of 2.0~W, the highest reported at this spectral range~\cite{Bernier2024}. Their design, based on a heavily erbium-doped fluoride fiber cavity terminated by photo-inscribed fiber Bragg gratings, achieved a slope efficiency of 46.5\%, highlighting the scalability and long-term stability achievable with monolithic mid-IR fiber architectures. Furthermore, a recent demonstration from the same international collaboration showed efficient supercontinuum generation in tantalum–gallate glass waveguides, with potential for applications in the mid-infrared spectral region \cite{Guerineau2025}.

One major obstacle in producing hybrid fiber–waveguide architectures has been the incompatibility between standard integrated photonic platforms and the mid-IR operational range of fluoride fibers. Silica, the dominant material in photonic integration, is opaque beyond 2.4 {\textmu}m, while chalcogenide glasses, though mid-IR transparent, suffer from poor thermal and mechanical stability, limiting their use in active laser platforms. While earlier mid-IR glass processing techniques were unable to achieve low-loss, three-dimensional waveguides with sufficient index contrast for compact functional devices, recent advances in year 2024~\cite{FernandezOE24} have demonstrated a solution—achieving a high refractive index change (~$\approx 2.25 \times 10^{-2}$) through bulk material modification—enabling the fabrication of integrated components such as directional couplers and on-chip cavities.

In this work, we demonstrate, for the first time, a fully integrated mid-IR fiber ring laser based on a fluoride glass photonic chip. Using femtosecond laser direct writing (FLDW), we inscribe directional couplers directly into bulk fluoride glass with an optimized composition, achieving high positive refractive index modifications suitable for guiding light at $\sim$2.8 {\textmu}m. The couplers are engineered with varying interaction lengths and coupling gaps to tailor their power splitting ratios.

A directional coupler with a 7 {\textmu}m gap and 1.5 mm interaction length was used to construct a compact ring resonator cavity, where it functions both as an intracavity feedback element and output coupler. The coupler’s 80:20 splitting ratio allows efficient out-coupling while maintaining sufficient feedback for lasing. When coupled to a cladding-pumped Er:ZBLAN fiber, the system generates stable lasing at 2716.7~nm with 14.2 mW of output power, demonstrating the viability of hybrid integration between fiber and chip in the mid-IR.

This approach marks a critical advance in mid-IR photonics by bridging the gap between bulk glass processing and fiber-based architectures. It enables the realization of compact, robust, and alignment-free mid-IR laser systems with embedded functionality such as power splitting, filtering, and beam shaping. Furthermore, the 3D writing capability of FLDW opens up possibilities for novel geometries and integrated sensing schemes that were previously unattainable with purely fiber-based designs.

\section{Materials and Methods}

The substrate used in this study is a custom fluoride glass composition known to support high positive refractive index changes upon femtosecond laser irradiation. A detailed report on the composition and its response to femtosecond laser pulses are provided in ref~\cite{FernandezOE24}. Notably, this composition is lanthanum-free, a characteristic that enhances ion migration under laser exposure and facilitates the formation of index changes exceeding $2 \times 10^{-2}$. This makes the glass particularly suitable for femtosecond laser direct writing of low-loss waveguides and directional couplers for mid-infrared applications.

Waveguide and coupler inscription was performed using a Pharos femtosecond laser system (Light Conversion) operating at a central wavelength of 1030 nm with a pulse duration of 240 fs and a repetition rate of 50 kHz. The laser beam was circularly polarized and focused into the bulk of the fluoride glass using a 40× objective lens (Olympus LUCPlan FL N, 0.6 NA). A controlled amount of spherical aberration was introduced by detuning the coverslip compensation collar of the objective lens by 500~{\textmu}m. This technique has been shown to significantly enhance the waveguide quality and index contrast in glass hosts~\cite{Song2011,Fernandez2015,Fernandez2022}.

Directional couplers were fabricated with varying interaction lengths ranging from 0 to 4.5 mm and three different waveguide separations of 5, 7, and 9 {\textmu}m at the coupling region. Each directional coupler consisted of two waveguides brought close together over the coupling length, with input and output ports separated by 350~µm. To maintain a uniform total device length of 16 mm across all couplers, the input and output straight sections of the waveguides were adjusted accordingly. The curved sections connecting the coupling region to the straight input/output arms were designed to span over 4 mm (radius of curvature of 18 mm), ensuring minimal bend loss while accommodating the necessary geometry for a compact layout.

The repetition rate (50 kHz), laser power (40 mW), sample translation speed (40~{\textmu}m/s), focussing objective aberration collar (500 {\textmu}m) and laser polarization (circular) were kept constant during all fabrications to ensure consistency in waveguide performance. All structures were written approximately 170 $\mu$m below the surface of the glass to minimize surface scattering and environmental contamination. Post-fabrication, the end facets of the samples were polished to enable efficient coupling to mid-infrared fibers for characterization and laser integration.

\section{Results and discussion}\label{sec4}
\subsection{Directional coupler characterization}
\begin{figure}[t]
    \centering

     \includegraphics[trim={0 0 0 0},width=12 cm]{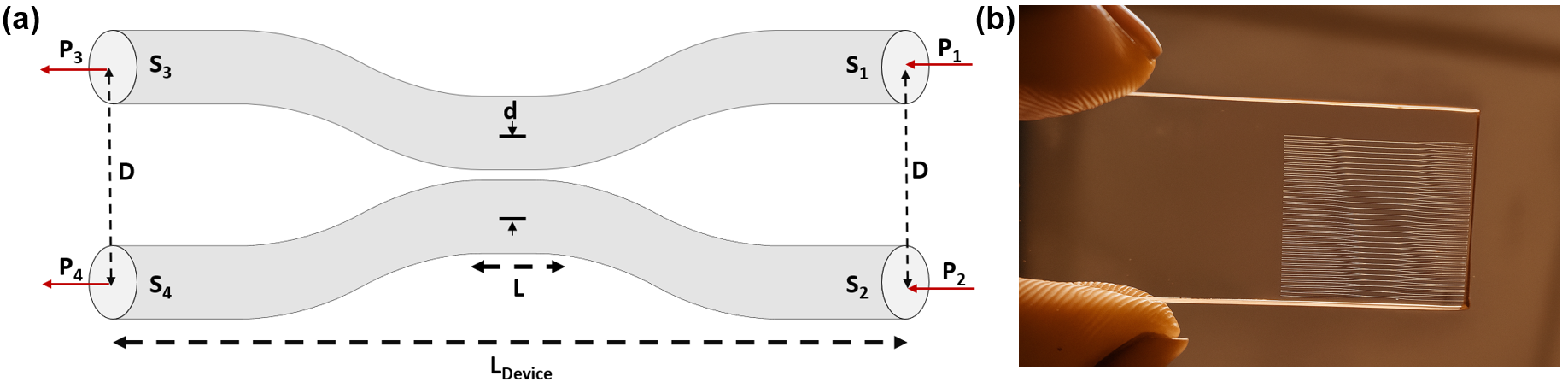}
        \caption{(a) Schematic illustration of the directional four-port ($S_{1}, S_{2}, S_{3}, S_{4}$) coupler. The input power $P_{1}$ or the input power $P_{2}$ are split between the output port 3 and the output port 4, according to the different coupling length L and coupling gap, d. (b) Photograph of the directional couplers inscribed into the ZBLAN glass with varying d and L. }
    \label{Coupler_Schematic}
    \end{figure}

The primary parameters controlling the coupler performance are the coupling length L and the coupling gap d as shown in figure~\ref{Coupler_Schematic}(a). A total of 30 four-port directional couplers are designed and fabricated, with the free parameter D uniformly set to 350 {\textmu}m to ensure compatibility with optical fibers of diameters larger than 250 {\textmu}m. All directional couplers shared a uniform total length L$_{Device}$ of 16 mm. Sketch of one directional four-port coupler with all the parameters labelled is shown in figure~\ref{Coupler_Schematic}(a). Figure~\ref{Coupler_Schematic}(b) shows the photograph of all the directional couplers after laser inscription but before cutting and polishing. The variation in coupling length L from 0 to 4.5 mm in 0.5 mm steps is clearly visible in the photograph.

We fabricated a side-pump coupler using the silica fiber taper-winding method ~\cite{S2020Fuseless,2024High} and pumped a Er\textsuperscript{3+}-doped ZBLAN fiber (the cladding is 250 $\mu$m and the core is 15 {\textmu}m) with a 976 nm laser source, thereby realizing a 2.8 {\textmu}m single-mode laser. A maximum output power of 2 W is obtained from the fiber laser. This laser was used as the source to characterize all the directional couplers in the glass.

\begin{figure}[t]
    \centering
    \includegraphics[trim={0 0 0 0},width=12 cm]{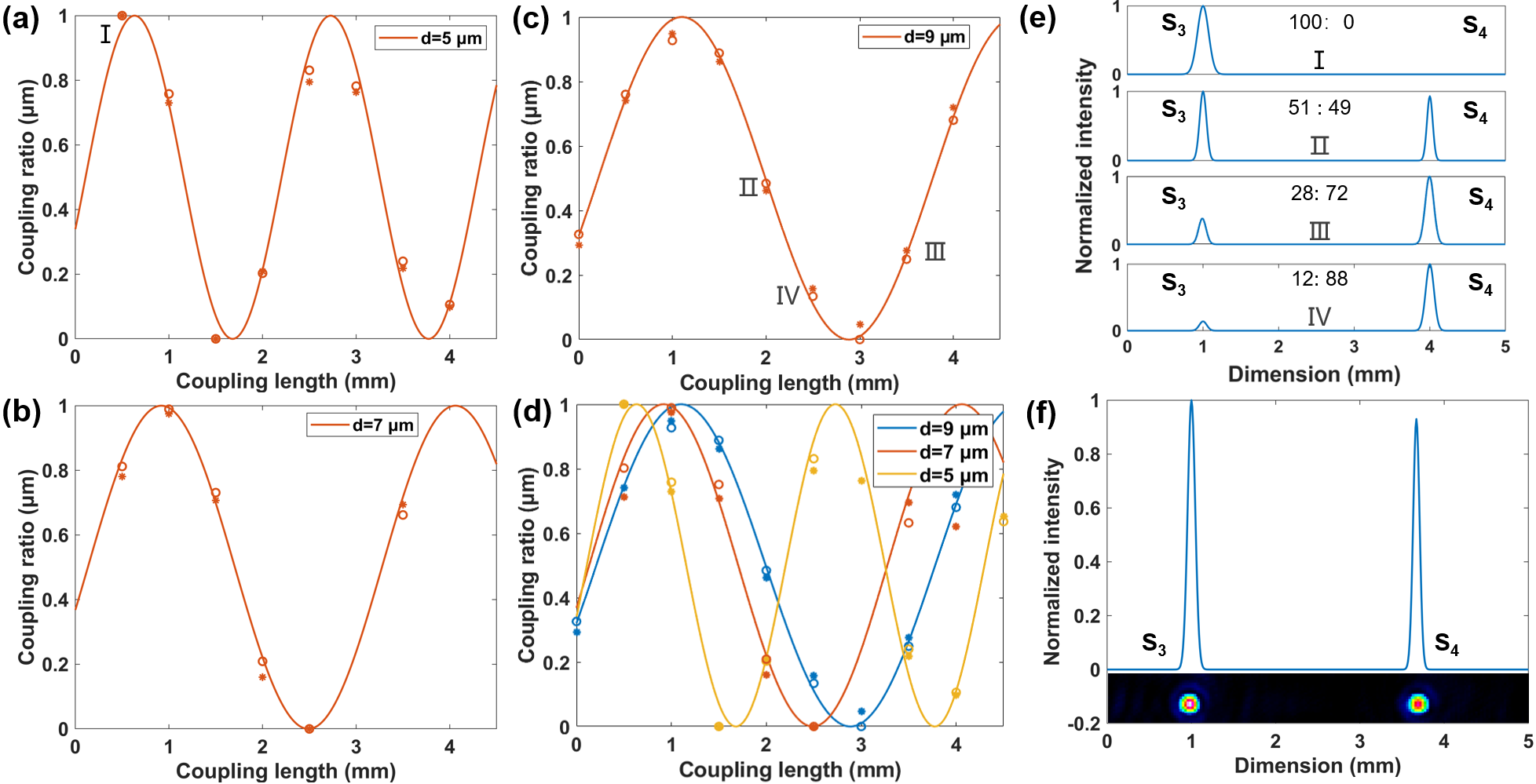}  
        \caption{(a) The relationship between the coupling length L and the coupling ratio for directional couplers with coupler spacing d of 5 {\textmu}m. (b) The relationship between the coupling length L and the coupling ratio for directional couplers with coupler spacing d of 7 {\textmu}m. (c) The relationship between the coupling length L and the coupling ratio for directional couplers with coupler spacing d of 9 {\textmu}m. (d) represents the consolidation of (a), (b) and (c). (e) Different coupling ratios ranging from 100:0 to 12:88 selected from a constant coupler spacing of 9 {\textmu}m with different coupling lengths. The laser light is coupled into the directional coupler through port $S_{1}$. In the figure, Arabic numerals identical to those in (a) and (c) correspond to the same waveguide. (f) Images of the output optical mode patterns from a directional coupler with a coupling length of 1 mm and a coupling spacing of 9 {\textmu}m.}
    \label{Coupler_Charact}
    \end{figure}

The coupling ratios achieved for different coupling lengths L and waveguide separations in the coupling region d are shown in figure~\ref{Coupler_Charact} (a). When laser light is coupled into the directional coupler via input port 1 or input port 2, the coupling ratios can be expressed as:

\begin{equation}
\eta_1 = \frac{P_1(L)}{P_0} \quad \text{and} \quad \eta_2 = \frac{P_2(L)}{P_0}
\end{equation}

Considering fabrication or measurement errors such as a small length offset $\Delta L$, the model is often fitted using:
\begin{equation}
\eta = \sin^2[\kappa(L + \Delta L)] = \sin^2(\kappa L + \phi)
\end{equation}

where $\phi$ is a phase offset used for empirical fitting. The two input ports of the directional coupler were repeatedly measured, revealing variation of less than $\pm 4.9\%$ (in the normalized coupling ratio), thus indicating comparable loss between the ports and highly symmetric coupling. Under the condition of a coupling spacing d = 9 {\textmu}m and variable coupling lengths L, the transverse cross-sectional mode profiles for different coupling ratios ranging from 100:0 to 12:88 are shown in figure~\ref{Coupler_Charact}(e). Furthermore, the beat length increases with coupler spacing (d). The beat length is identified as the distance along the coupler at which the power distribution between the two arms of the coupler completes a full transfer cycle. Experimentally, it is obtained from the spacing between successive maxima (or minima) in the measured output power oscillations as a function of coupling length.

The two-dimensional intensity distribution and corresponding beam spot pattern output from a directional coupler with a coupling length of 1 mm and a coupling spacing of 9 $\mu$m, as shown in figure~\ref{Coupler_Charact}(f). The beam profiles were recorded by a mid-infrared camera at identical positions. The directional couplers presented in this work were designed primarily to demonstrate coupling functionality and were not optimized with respect to propagation loss. The bend radius employed in this study was approximately 18.5 mm, among the lowest reported for this platform. In addition, the waveguide width strongly influences the wavelength of operation, and further optimization of this parameter—together with bend radius and the ideal fluence required to achieve ultra-low-loss bends—warrants a dedicated investigation, which lies beyond the scope of this paper.

\begin{figure}[t]
    \centering
    \includegraphics[trim={0 0 0 0},width=8 cm]{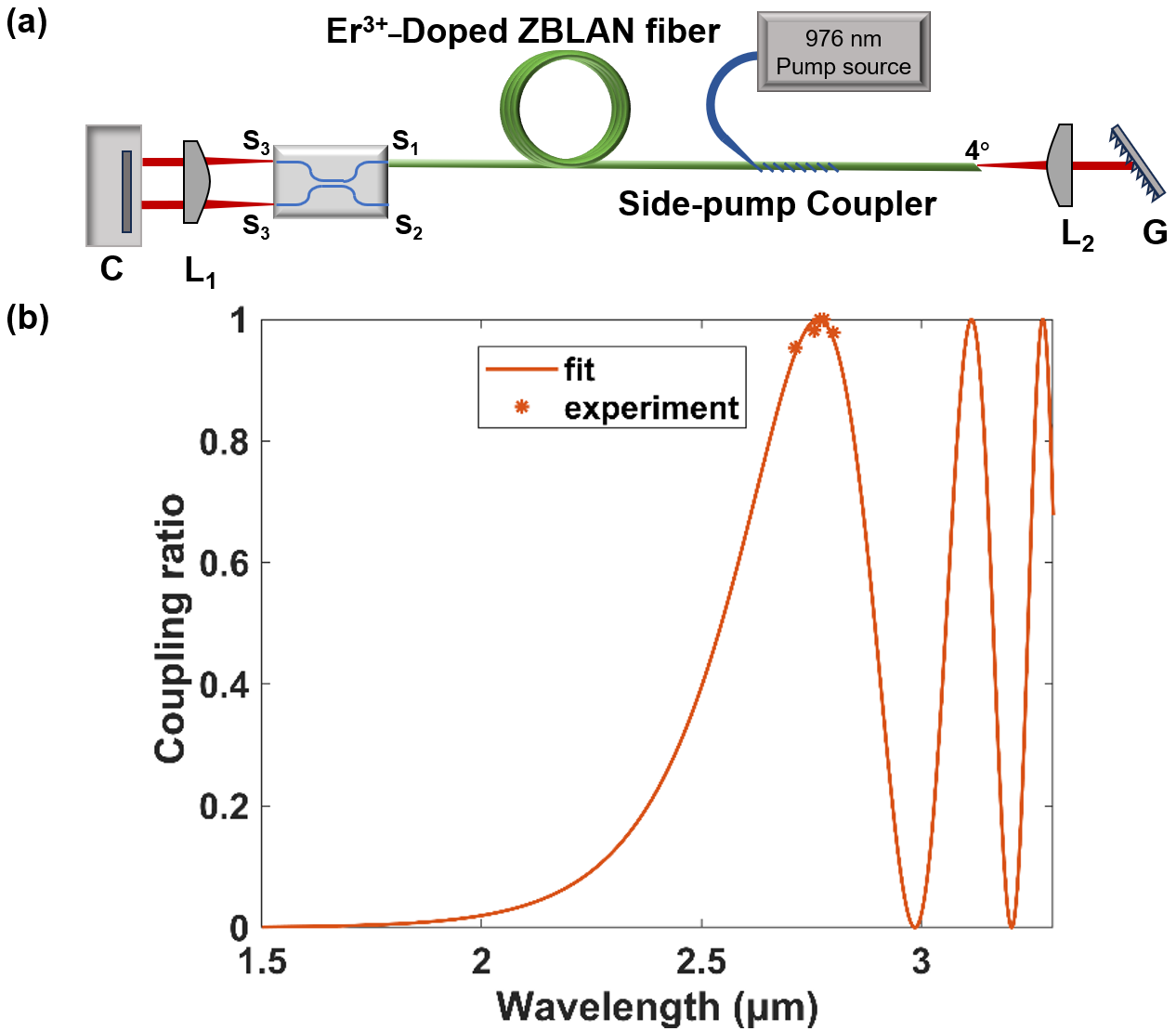}  
        \caption{Sketch of the structural setup for a mid-infrared wavelength-tunable fiber laser. The plano-convex CaF\textsubscript{2} lenses L\textsubscript{1} and L\textsubscript{2} have focal lengths of 20 mm. The reflective diffraction grating G has a center wavelength of 3.1 {\textmu}m. (b) The relationship between the wavelength and the coupling ratio for directional couplers with coupler spacing d of 7 {\textmu}m and coupling length L of 1 mm.}
    \label{Simulation}
    \end{figure}

A mid-infrared reflective diffraction grating (Thorlabs GR1325-45031) was employed to construct a wavelength-tunable fiber laser, as shown in figure~\ref{Simulation} (a). One end-face of the ZBLAN fiber is angle-cleaved at 4$^{\circ}$. This effectively configures the laser resonator to operate via feedback from the output end-face and grating reflection, rather than relying on Fresnel reflections from both ZBLAN fiber end-faces. Wavelength tuning across 2710–2780 nm is achieved by angular adjustment of the reflective diffraction grating. The wavelength-tunable fiber laser operates at 100 mW output power. Equation (1) reveals the functional relationship between the coupling ratio and the product of the coupling coefficient $\kappa$ and coupling length L. When the coupling length is held constant and the coupling coefficient varies as a function of wavelength, the functional relationship may be expressed as:

\begin{equation}
\eta_1(\lambda) = \sin^2\left(\kappa(\lambda)(L + \Delta L)\right)
\end{equation}

This equation was used to model the spectral dependence of the directional coupler's coupling efficiency. The fitting approach enabled accurate representation of the wavelength-dependent behavior, as shown in figure~\ref{Simulation}(b).

\subsection{Integrated ring cavity laser}

We present the design of a ring-cavity mid-infrared fiber laser. The ring-cavity fiber laser comprises an on-chip directional coupler, a side-pumping coupler, passive ZBLAN fiber, and erbium-doped ({Er\textsuperscript{3+}) gain ZBLAN fiber. The laser, pumped at 976 nm, achieved continuous-wave operation at 2.7 $\mu$m. The directional coupler selected for the circulator features design parameters of a 1.5 mm coupling length and a 7-{\textmu}m coupler separation. The simulated power-splitting ratio is shown in Fig.~\ref{Coupler_Simulation}.

\begin{figure}[t]
    \centering
    \includegraphics[trim={0 0 0 0},width=6.5 cm]{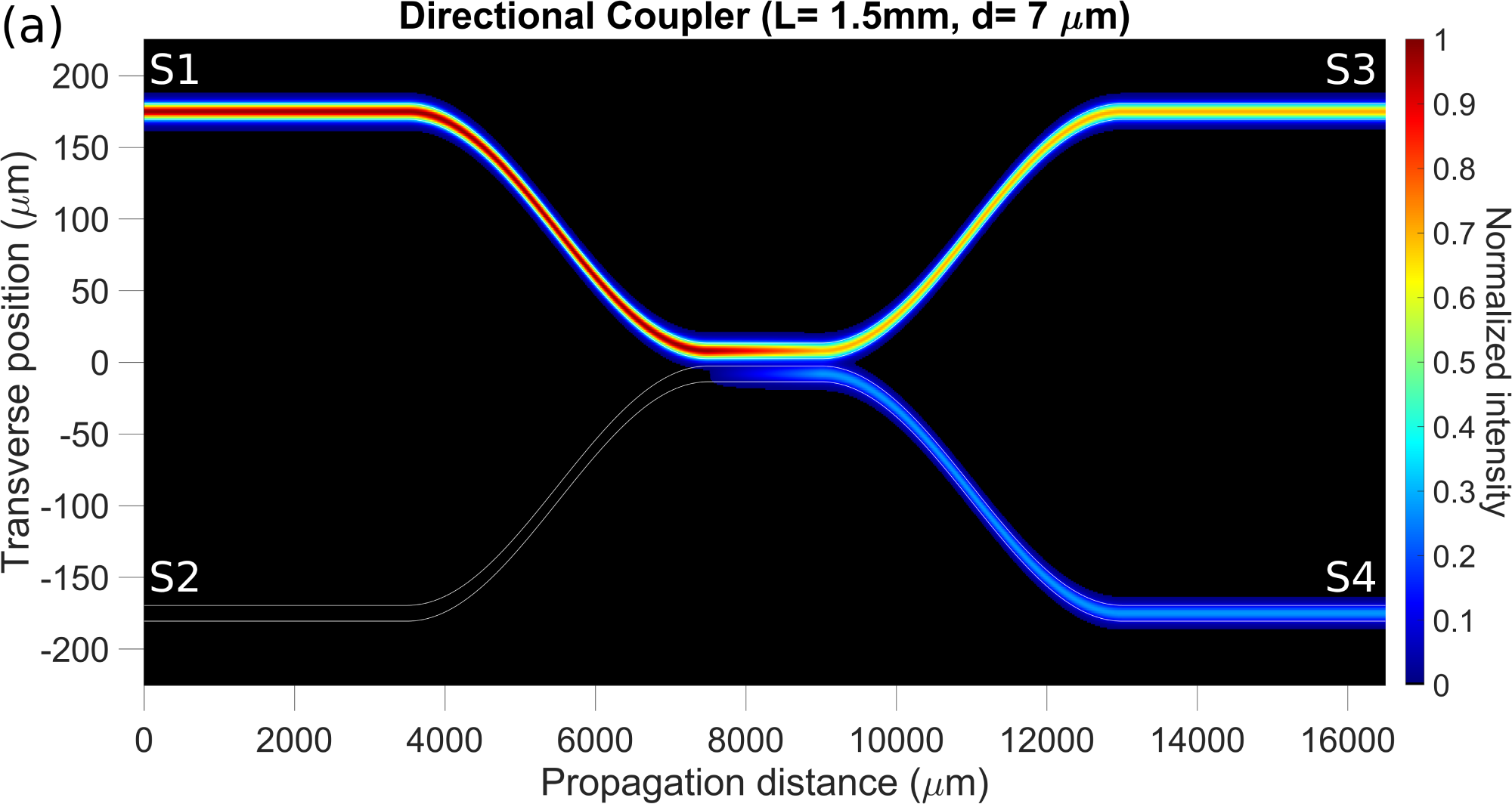} 
    \includegraphics[trim={0 0 0 0},width=6.2 cm]{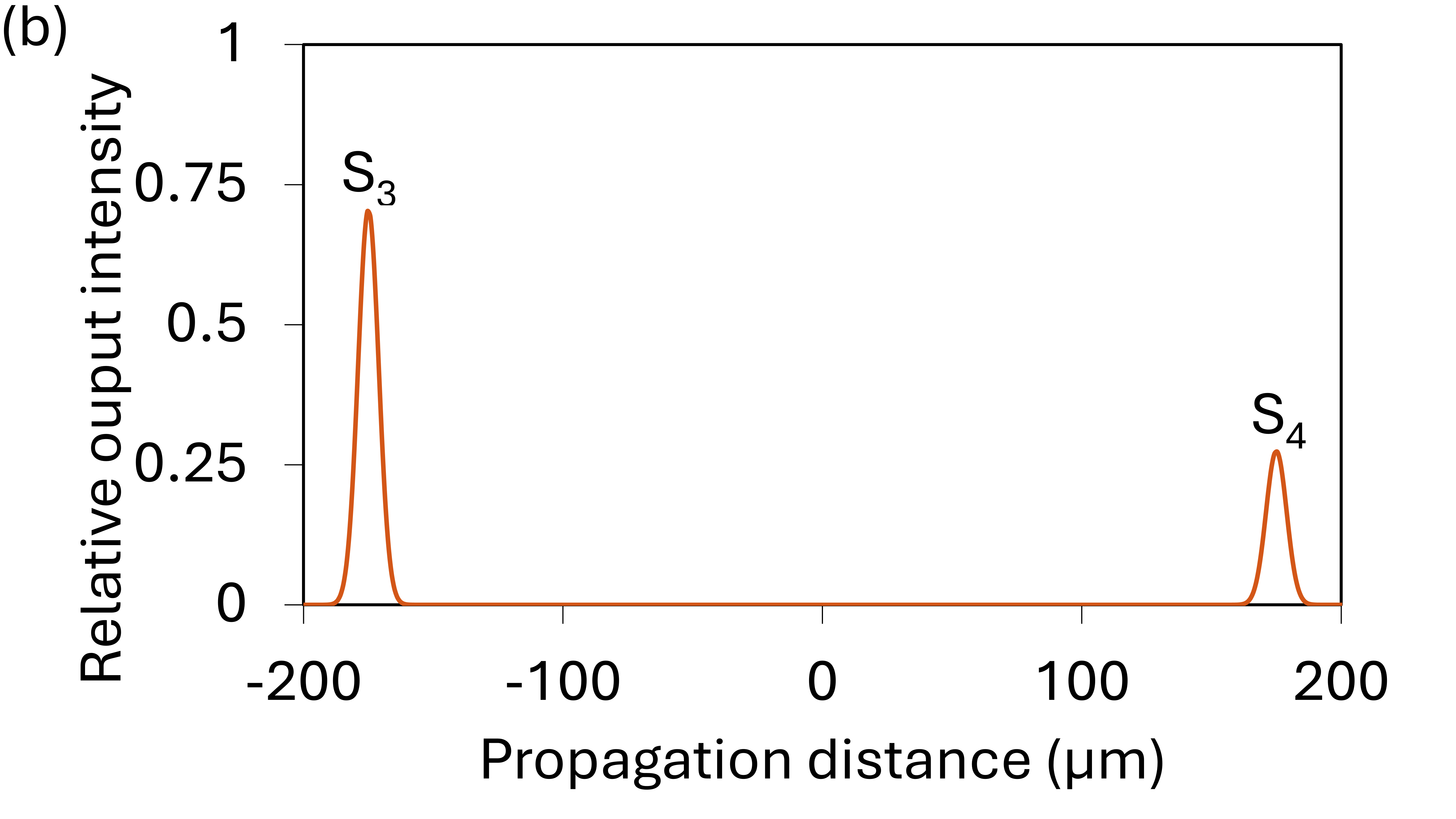} 
        \caption{(a) Simulated mode propagation intensity in the directional coupler at 2720 nm. (b) Corresponding horizontal line profiles at the output ports (S3 and S4) when excitation is applied at S1. }
    \label{Coupler_Simulation}
    \end{figure}

Figure~\ref{Ring_laser} (a) shows the schematic diagram of the ring-cavity fiber laser. The gain fiber is directly butt-coupled to the ports of the on-chip directional coupler. The gain fiber is a 2.5-m-long, double-clad, Er\textsuperscript{3+}-doped ZBLAN fiber with a dopant concentration of 7 mol\% (Le Verre Fluoré, France). It possesses a core diameter of 15 {\textmu}m, a numerical aperture (NA) of 0.12, and incorporates a double-D-shaped cladding with cross-sectional dimensions of 240 x 260 {\textmu}m. This cladding is coated with a low-refractive-index polymer, which functions as the secondary cladding.

The side-pumping coupler employs a taper-and-wrap technique to couple the 976-nm laser diode pump radiation into the inner cladding of the gain fiber. The ring-cavity fiber laser achieved a maximum single-port output power of 14.2 mW. Figure~\ref{Ring_laser} (b) shows the output spectrum; continuous-wave laser output is generated at 2716.7 $\mu$m and its mode profile is also shown in the inset of figure~\ref{Ring_laser} (b).

\begin{figure}[h]
  \centering
  \includegraphics[trim={0 0 0 0}, width=1 \textwidth]{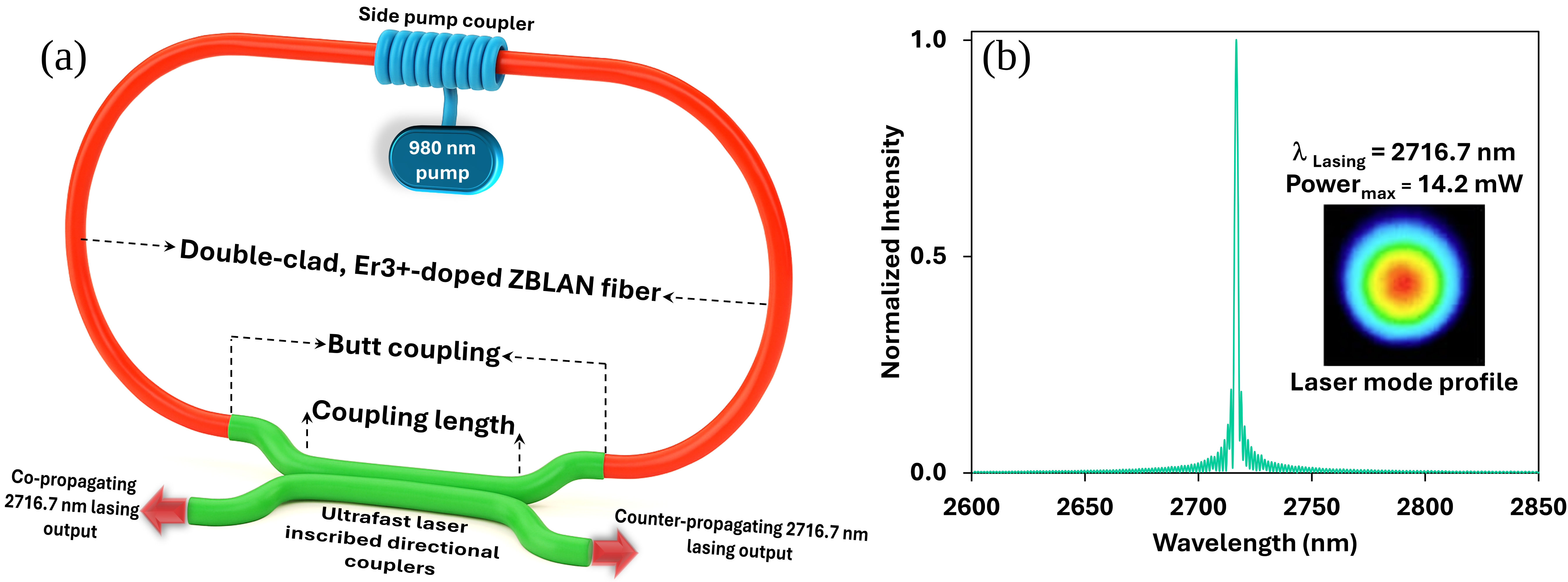}
  \caption{(a) Schematic of the mid-IR ring-cavity fiber laser configuration. (b) The laser spectrum at 2716.7 nm and also the mode profile of the laser at the coupler output.}
  \label{Ring_laser}
\end{figure}

\section{Conclusion}\label{sec13}
To summarize, this study demonstrates the development of a compact mid-infrared ring laser utilizing on-chip directional couplers fabricated in fluoride glass platforms. The couplers were fabricated via femtosecond laser direct inscription techniques into fluoride glass, creating low-loss waveguides characterized by a substantial positive refractive index modification on the order of 10\textsuperscript{-2}. The waveguides were designed to efficiently support light propagation at a wavelength of 2.8 {\textmu}m. A set of directional couplers with different arm spacings and coupling lengths were produced and evaluated, demonstrating power division ratios spanning from 100:0 to 50:50. Specifically, a coupler featuring a 7 {\textmu}m spacing and a 1.5 mm interaction length—achieving an 80:20 power-splitting ratio—was combined with a cladding-pumped Er:ZBLAN fiber to assemble a ring laser resonator. The integrated system generated stable laser emission at 2716.7 nm wavelength, yielding an output power of 15 mW. This achievement marks the first demonstration of a fully integrated fiber-to-chip mid-infrared laser system and introduces novel opportunities for three-dimensional photonic integration within fluoride glass materials, offering design capabilities that transcend the limitations of conventional fiber-based configurations.

\backmatter

\bibliography{References}

\end{document}